# Soft Carrier Multiplication by Hot Electrons in Graphene

Anuj Girdhar[1,3] and J.P. Leburton[1,2,3]

1) Department of Physics
2) Department of Electrical and Computer Engineering, and
3) Beckman Institute
University of Illinois at Urbana-Champaign
Urbana, Il 61801

## Abstract

By using Boltzmann formalism, we show that carrier multiplication by impact ionization can take place at relatively low electric fields during electronic transport in graphene. Because of the absence of energy gap, this effect is not characterized by a field threshold unlike in conventional semiconductors, but is a quadratic function of the electric field. We also show that the resulting current is an increasing function of the electronic temperature, but decreases with increasing carrier concentration.

In the last decade, owing to its remarkable physical properties, graphene has emerged as a new material for high performance electronic applications [1,2,3,4]. Its strictly two-dimensional (2D) nature consisting of a single sheet of atoms combined with the energy dispersion $E = \pm\hbar v_F k$, linear in the electron momentum $k = \sqrt{k_x^2 + k_y^2}$, which results in the same velocity $v_F \sim 10^8$ cm/s for all carriers, is well suited for high speed electronics with planar technology [5,6,7]. Another important feature of graphene is the reduction of energy gap to a single (Dirac) point in k-space, which favors interband processes such as impact ionization or Auger recombination, compared to conventional semiconductors [5]. Recently, the scattering rates for impact ionization and Auger recombination in graphene have been derived for inverted electron-hole populations, for which the scattering times were found longer than 1ps for concentrations less than $10^{12}$/cm$^2$ [8]. This situation is, however, different from electronic transport, where disturbances from equilibrium arises from the external electric field, creating the conditions for interband interaction, which affects the overall transport in graphene at relatively low fields.

In this letter we consider the effect of carrier multiplication induced by interband inter-carrier interaction caused by hot electrons transport in mono-layer graphene. We assume carrier populations are away from the intrinsic regime, and the graphene layer is mostly n-type (because of electron-hole symmetry, the p-type case is similar), as achieved in electronic devices.

In these conditions, impact ionization with carrier multiplication occurs when a conduction band electron (**1**) scatters with another electron in the valence band (**2**), resulting in two secondary electrons in the conduction band (**1'** and **2'**), creating a hole in the valence band (Fig. 1). In this inter-electron collision, both momentum and energy are conserved, which around the Dirac point for intra-valley processes yields

$$\boldsymbol{k_1} + \boldsymbol{k_2} = \boldsymbol{k_{1'}} + \boldsymbol{k_{2'}}, \tag{1}$$

$$k_1 - k_2 = k_{1'} + k_{2'} \tag{2}$$

thereby accounting from the fact that particle 2 is below the gap, so $E_2 = \hbar v_F k_2 < 0$, and dropping the $\hbar, v_F$ constant factors in these equations. Squaring both equations and taking the difference yields a condition for the relative orientations of the initial and final wavevectors, namely that they all lie on a line. Wavevectors $\boldsymbol{k_1}, \boldsymbol{k_{1'}}$, and $\boldsymbol{k_{2'}}$ must all be parallel while $\boldsymbol{k_2}$ is antiparallel. Here, only intra-valley scattering events are considered, since the analysis of the scattering matrix element reveals that large momentum-transfer interactions are significantly suppressed.

By Fermi's golden rule, the transition probability per unit time is given by

$$S(\boldsymbol{k_1}, \boldsymbol{k_{1'}}; \boldsymbol{k_2}, \boldsymbol{k_{2'}}) = {2\pi}/{\hbar} \, |M|^2 \delta(E_1 + E_2 - E_{1'} - E_{2'}), \tag{3}$$

$$M = {e^2}/{2\kappa_{eff} \kappa_0 q \varepsilon(q)}, \tag{4}$$

where $M$ is the scattering matrix element of the screened Coulomb interaction between electrons including the direct and exchange interactions [8,9]. $\kappa_{eff}$ is the effective dielectric constant of single-layer graphene on an SiO$_2$ substrate, and $\varepsilon(q)$ is the static screening dielectric function where $q$ is $|\boldsymbol{k_1} - \boldsymbol{k_{1'}}|$,

$$\varepsilon(q) = 1 + {e^2 k_F}/{\pi \hbar v_F \kappa_s \kappa_0 q} \tag{5}$$

as calculated by Hwang *et al.* using the random phase approximation, which combines inter and intraband scattering events by use of an effective permittivity which takes both processes into account [10]. Here, $\kappa_s$ is the background dielectric constant due to the substrate.

In order to calculate the total scattering rate as a function of electric fields, we assume the electron concentration in the conduction band is high enough, and the fields small enough to justify an electronic temperature model [11]. In these conditions, the distribution function reads

$$f(\boldsymbol{k}) = \left[\exp\left(\frac{\hbar v_F(|\boldsymbol{k} - \boldsymbol{k_D}| - k_F)}{k_B T_e}\right)\right]^{-1} \tag{6}$$

with $\boldsymbol{k_D} = {eF\tau}/{\hbar}$, so $f(\boldsymbol{k})$ is a Fermi-Dirac distribution function displaced in the electric fields with quasi-Fermi level and electronic temperature $T_e$. Here we have assumed the field points in the positive $x$ direction. $\tau$ is the mean free time as a result of collisions with impurities and phonons [12].

The scattering rate, or the inverse scattering time, is given by

$$\tau_{ee}^{-1} = \sum_{k_2, k_{1'}, k_{2'}} S(\boldsymbol{k_1}, \boldsymbol{k_{1'}}; \boldsymbol{k_2}, \boldsymbol{k_{2'}}) \tag{7}$$

$$\times [f(\boldsymbol{k_2})][1 - f(\boldsymbol{k_{1'}})][1 - f(\boldsymbol{k_{2'}})]\delta_{k_1+k_2, k_{1'}+k_{2'}}$$

where S is the probability per unit time defined in eq. 3 and the sum is over all wavevectors **1'**, **2'**, and **2** [13]. The exclusion principle is enforced by using the rejection technique after assuming high carrier densities and therefore significant carrier degeneracy. We use a Gauss-Kronrod quadrature routine that assumes a linear energy dispersion and includes band degeneracy to compute the integral in eq. 8 [14]. Here we mention the nontrivial transformation of the δ-functions for energy and momentum conservation in the transition probability,

$$\sum_{\mathbf{k_2}, \mathbf{k_{1'}}, \mathbf{k_{2'}}} \delta(k_1 + k_2 - k_{1'} - k_{2'}) \delta_{k_1+k_2, k_{1'}+k_{2'}} \quad (8)$$

$$\Leftrightarrow \frac{A^2}{(2\pi)^3} \iint dk_{1'} d\theta_{1'} \iint dk_2 d\theta_2 \frac{k_{1'} k_2 (k_1 - k_2)}{\sqrt{k_{1'} k_{2'} k_1 k_2}} \frac{\delta(\theta_{1'} - \theta_{2'}) \delta(\theta_1 - \theta_2)}{|\cos[\frac{1}{2}(\theta_{1'} - \theta_{2'})]||\sin[\frac{1}{2}(\theta_1 - \theta_2)]|}$$

where $\theta_{1'}$ ($\theta_{2'}$) and $\theta_1$ ($\theta_2$) are the $\mathbf{k_{1'}}$ ($\mathbf{k_{2'}}$) and $\mathbf{k_1}$ ($\mathbf{k_2}$) wavevector angles with respect to the field direction.

In fig. 2 we show the scattering rate as a function of the initial carrier energy for an electronic temperature $T_e$=1200K and a field of 1kV/cm, which reaches approximately 0.1 ps$^{-1}$ for carrier energies near ~0.2 eV at a carrier density of $1\times10^{12}$ /cm$^2$. The rate decreases by several orders of magnitude for lower energies and higher concentrations. Such strong decrease in the scattering rate is due to the high occupation of low energy states far below the quasi-Fermi level. Indeed, the most favorable states for secondary electrons lie in the conduction band at mid-way between $k_1$ and $k_2$, which to be empty requires large incident energy $E_1$, and $E_2$ close to the Dirac point. The scattering rate is about .2 ps$^{-1}$ for a carrier energy of ~.5 eV and a carrier concentration of $5\times10^{12}$cm$^{-2}$, increasing by many orders of magnitude for lower energies. These rates are much smaller than those from electron-electron scattering events in the conduction band only (~100 ps$^{-1}$ at .5 eV for a carrier density of $5\times10^{12}$/cm$^2$), as calculated by Li *et al*, which are due to is the relative availability of states below the Fermi level in intraband scattering events [15].

The inset shows the scattering rate variation for different electronic temperature, and for a population of $5\times10^{12}$ cm$^{-2}$ (with quasi-Fermi level $E_F$=.26 eV). The rate increases rapidly with temperature for incident electrons near the quasi-Fermi level, spanning several orders of magnitude between electron temperatures of 300K and 1200K. This is due to the increase of available states below the quasi-Fermi level as a result of thermal broadening of the distribution. At incident particle energies much higher than the quasi-Fermi level, the scattering rates become independent of temperature since the secondary electrons at mid-way between $k_1$ and $k_2$ will always find empty states.

The net generation rate per unit area resulting from impact ionization is given by

$$U = -\frac{8}{A}\sum_{\substack{\mathbf{k}_1,\mathbf{k}_{1'}\\ \mathbf{k}_2,\mathbf{k}_{2'}}} S \times \left\{ \begin{array}{l} [f(\mathbf{k}_{1'})][f(\mathbf{k}_{2'})][1-f(\mathbf{k}_1)][1-f(\mathbf{k}_2)] - \\ {[f(\mathbf{k}_1)][f(\mathbf{k}_2)][1-f(\mathbf{k}_{1'})][1-f(\mathbf{k}_{2'})]} \end{array} \right\} \quad (9)$$

where the factor 8 accounts for spin and valley degeneracies, and electron-hole multiplication. Here, the first term corresponds to impact ionization (generation), while the second is the Auger term (recombination), non-negligible at low fields.

Figure 3 shows the impact ionization net generation rate as a function of electric fields for three different electronic temperatures. One notices that the generation rate varies by several orders of magnitude for an electronic temperature variation of 300K to 1200K. Again, this is due to the thermal broadening of the distribution, which uncovers available states for secondary electrons just above the Dirac point. Unlike in conventional semiconductors that are characterized by exponential increase of the current, our model shows a power law for the current variation with electric fields that becomes quadratic for low fields. This soft carrier multiplication process is due to the fact that the leading term in the integral in eq. 9 is quadratic in the electric fields. Indeed, using the low-field approximation for the distribution one can be split into an even zero-field component and an odd term proportional to the field $F$

$$f(\mathbf{k}) = f_0(\mathbf{k}) + f_1(\mathbf{k}) \qquad (10)$$

$$f_1(\mathbf{k}) = -\frac{eF\tau}{\hbar} \equiv f_0(\mathbf{k}) = -\frac{eF\tau}{\hbar} \cos\theta \frac{\partial}{\partial k} f_0(\mathbf{k}) \qquad (11)$$

Then, the product of the distribution functions can be expressed as a sum of powers of the fields, the odd terms vanish once the angular $\mathbf{k_1}$ integral is performed since they are also odd in powers of $\cos\theta$. The zero-order term vanishes due to detailed balance, which results in a quadratic dependence of the current density on applied field for small fields. However, the exponent of the power law decreases with increasing field as a result of the sign of contributing higher-order terms in the expansion of the distribution function. This power law appears to be in good agreement with the model in ref. [16] that uses the quadratic term for the upkick current at low biases in G-FET devices.

Typical impact ionization and Auger recombination rates for hot electrons are strongly dependent on temperature and range from $\sim 10^{17}$ s$^{-1}$cm$^{-2}$ to $\sim 10^{23}$ s$^{-1}$cm$^{-2}$ for temperatures varying between 300K and 1200K at a field of 10 kV/cm. When compared to values obtained by Meric *et al* for a G-FET device, we obtain a corresponding current of ~0.3 mA comparable to the experimental data, by assuming that impact ionization takes place over 0.1um range on the drain side for electronic temperatures of 1200K and carrier densities of ~1-2×10$^{12}$ cm$^{-2}$, suggesting that carrier multiplication and Auger recombination events are important processes in such devices [17].

In conclusion, we proposed a model for hot carrier impact ionization and Auger recombination depending on the electronic temperatures and carrier concentrations in graphene. These processes are strongly dependent on the availability of states near the Fermi-level, as shown by large variations in response to thermal broadening and shifting of the Fermi energy, which results in a power law of the multiplication current with electric field. The power law dependence of the current density on applied field is in agreement with current upkick observed in graphene FET devices at low gate bias.

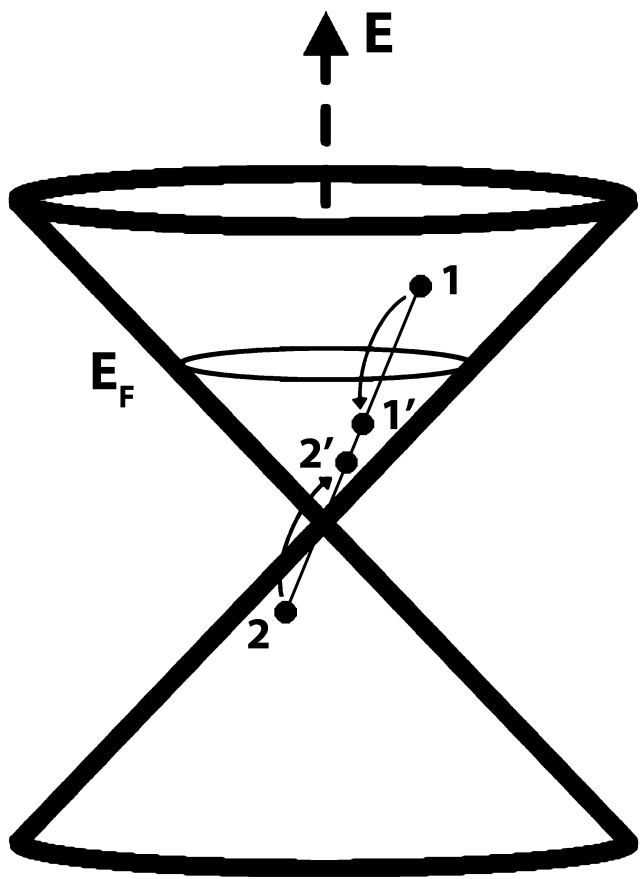

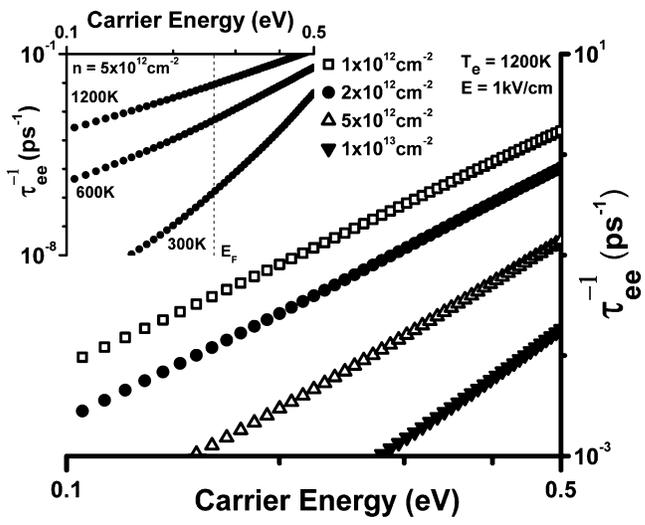

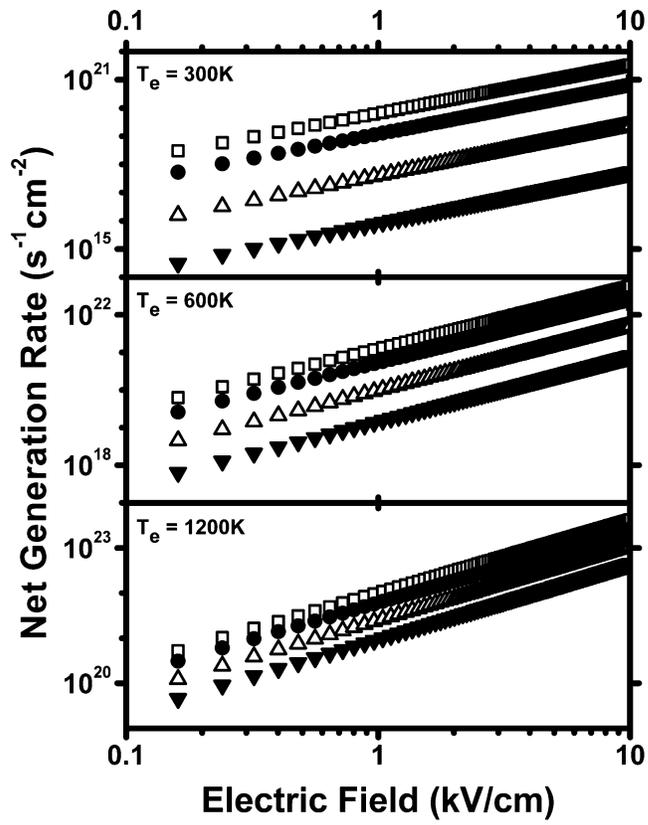

**Figure Captions:**

1. Schematic of a carrier multiplication scattering event in the Dirac cone in graphene. Conduction electron 1 scatters with valence electron 2, creating two conduction electrons 1' and 2' in addition to a valence hole. The reverse event is Auger recombination.
2. Electron-electron scattering rate for carrier multiplication as a function of the energy of the incident electron (particle 1) for various carrier concentrations at an electronic temperature of 1200 K, electric field of 1 kV/cm, and mean free time of 30 fs [17]. The inset shows the scattering rate for a carrier concentration of $5\times10^{12}$ /cm$^2$ for electron temperatures varying between 300K to 1200K.
3. Current density as a function of applied fields for temperatures 300K to 1200K at various carrier concentrations and mean free time of 30 fs [17]. The same symbols as figure 2 are used for different carrier concentrations.